\documentstyle[12pt]{article}
\newcommand{\be}{\begin{equation}}
\newcommand{\en}{\end{equation}}
\newcommand{\bea}{\begin{eqnarray}}
\newcommand{\ena}{\end{eqnarray}}

\newcommand{\hbo}{\hbox to 1 true cm {\hfill } }

\newcommand{\Tr}{\hbox{Tr}}

\def\dslash{\partial\kern-.5em\slash}
\def\kslash{k\kern-.5em\slash}

\begin{document}
\vglue 1truecm

\vbox{
  \hfill NBI-HE-93-56
}
\vbox{ \hfill UNIT\"U-THEP-11/1993 }
\vbox{ \hfill September 1993 }

\vfil
\centerline{\bf \large The non-trivial phase of $\phi ^{4}$-theory
in an external magnetic field }
\centerline{\bf \large and at finite temperature$^1$}

\bigskip
\medskip
\centerline{ K.\ Langfeld }
\bigskip
\medskip
\centerline{Niels Bohr Institute, University of Copenhagen,
   Blegdamsvej 17 }
\centerline{DK--2100 Copenhagen, Denmark }
\medskip
\centerline{and}
\medskip
\centerline{Institut f\"ur theoretische Physik, Universit\"at
   T\"ubingen}
\centerline{D--74076 T\"ubingen, Germany}
\bigskip

\vfil
\begin{abstract}

The effective potential for the order parameter $\phi ^{\dagger }
\phi$ is investigated in massless $\phi ^{4}$-theory in the presence
of magnetic fields at finite temperature. It is found that the first
order nature of the phase transition from the non-trivial to the
perturbative ground state at a critical temperature is unchanged by
the presence of magnetic fields, though they do increase the
critical temperature and weaken the barrier separating the
perturbative from the non-trivial phase.  The results might be relevant
for the electro-weak phase transition as well as for inflationary models
of the early universe.

\end{abstract}

\vfil
\hrule width 5truecm
\vskip .2truecm
\begin{quote}
$^1$ Supported by DFG under contract Re $856/1 \, - \, 1$
\end{quote}
\eject
{\it 1.\ Introduction \/ }

In many applications $\phi ^{4}$-theories are designed to describe
the dynamics of phase transitions. Even if the fundamental theory
consists of multiple degrees of freedom of different Lorentz
representations, it can be effectively described by a $\phi ^{4}$-theory
if the transition is driven by a single scalar degree of freedom.
Therefore it is important to treat the $\phi^{4}$-theory in an
renormalisation group invariant way, first, to ensure that no spurious
contributions are picked up by subtracting the divergencies, and, second,
to guarantee that the $\phi ^{4}$-theory decouples from the underlying
microscopic physics.


Models of the inflationary universe~\cite{gu80,li83,as82,pi84,hol84}
serve as an important example. Such models are most often described by
a phase transition which occurs at very early times whose dynamics
support a very rapid expansion. This expansion or inflation of the
universe simultaneously solves several cosmological
problems~\cite{ko90}.  Although the underlying theory might be some
supersymmetric or supergravity (see e.g., \cite{hol84}) Grand Unified
Theory (GUT), the phase transition is described in terms of a single
scalar particle (inflaton)~\cite{gu80,li83}, the other degrees of
freedom being regarded as spectators and the most successful models
rely on very flat effective potentials upon which the evolution of the
field is like classical slow rolling.  However, if such flat
potentials are to arise from perturbative quantum field theory, they
require extremely small scalar couplings in order for density
fluctuations, produced during inflation, to be small
enough~\cite{ko90}. Such models tend to suffer from very low re-heat
temperatures as a result of the weak coupling.  However, scalar $\phi
^{4}$-theory could still hold the key.  Recently~\cite{fo93}, it was
shown that a non-perturbative treatment of
$\phi^{4}$-theory~\cite{la1,la2,la3} can effectively provide a slow-roll
potential without relying on tiny couplings. In particular, it is
non-perturbative effects which generate the very flat part of the potential
rather than small couplings so that sufficiently large re-heat
temperatures need not be ruled out~\cite{fo93}.

The electro-weak phase transition is a second important example of a
transition driven by a scalar particle (Higgs). In contrast to the
inflaton, the Higgs-particle though its existence has not been
established, is assumed to be physical.

Both of the above examples rely on a non-trivial phase structure of
$\phi ^{4}$-theory. Non-perturbative $\phi ^{4}$-theory has been
investigated by several authors~\cite{ab76,ste84,ok87} and recently in
an approach which is explicitly renormalisation group
invariant~\cite{la1,la2,la3}. In the latter, the effective potential
for the scalar field $\phi $~\cite{la1} and the composite field
$\phi ^{2}$~\cite{la2,la3} was calculated for massless $\phi
^{4}$-theory and two phases were found.  In one phase the effective
potential for $\phi $ coincides with the one-loop renormalisation
group improved potential. However, this perturbative phase is unstable
since a second phase with a non-vanishing scalar condensate has lower
vacuum energy density.  Since the effective potential for $\phi $ is
convex in both phases (implying $\langle \phi \rangle =0$), the scalar
condensate $\phi ^{\dagger } \phi $ is the appropriate order parameter
rather than the field $\phi $ itself. The effective
potential~\cite{la2} and the effective action~\cite{fo93} in a
derivative expansion have also been investigated at finite
temperature. It was seen that the non-trivial vacuum undergoes a first
order phase transition to the perturbative ground state if the
temperature is large enough. The implications of such a phase
transition for the inflating universe were discussed in~\cite{fo93}.

As mentioned above, the scalar field is usually some effective
degree of freedom rather than a fundamental field. The
description of the phase transition in terms of a single degree
of freedom might be a good approximation if the scalar theory
is treated in a renormalisation group invariant manner
\footnote{ This ensures that the only connection to the underlying
microscopic theory is made by the renormalisation conditions.},
and if
the spectator fields do not change the nature of the phase transition.
In the context of the phase transition in the early universe, the
spectator fields might be some remaining gauge field
degrees of freedom of the fundamental Grand Unified Theory. In the
context of the electroweak phase transition it was recently observed
that the symmetry breakdown of the gauge invariance of the standard
model produces large magnetic fields~\cite{va91,en93}. These fields
imply a small residual magnetic field in the present epoch, which, if
amplified by a galactic dynamo mechanism (see e.g. \cite{ruz88}),
could accounts for the galactic fields observed today~\cite{en93}. One
might ask whether such large fields would qualitatively change the
phase transition.

In this letter we investigate the effective potential of the order
parameter $\phi ^{\dagger } \phi $ in massless, charged $\phi
^{4}$-theory in an external magnetic field at finite temperature. We
will find that the phase transition from the non-trivial vacuum to the
perturbative ground state at the critical temperature is still of
first order. Details of the effective potential are however affected,
the critical temperature is increased in the presence of a
magnetic field, and the barrier, separating the perturbative and the
non-trivial state, becomes smaller with increasing magnetic field. In
context of the electroweak phase transition, this might indicate that
the phase transition is hastened toward the end when large magnetic fields
are produced.

\medskip
{\it 2.\ The non-perturbative approach in the presence of magnetic fields }

We start with the Euclidean generating functional for scalar
Green's functions in the presence of a magnetic field, i.e.,
\bea
Z[j] & = & \int {\cal D} \phi \; {\cal D} \phi ^{\dagger } \;
\exp \bigl( - \int _{0} ^{1/T} d \tau \; \int d^{3}x \;
[ \frac{1}{2} (D_{\mu } \phi ) ^{\dagger } D_{\mu } \phi
+ \frac{1}{2} m^{2} \phi ^{\dagger } \phi
\nonumber \\
&+& \frac{ \lambda }{24}
(\phi ^{\dagger } \phi )^{2} + \frac{1}{ 4 e_{0} } F_{\mu \nu }
F_{\mu \nu } - j(x) \phi ^{\dagger } \phi ] \bigr)
\label{eq:1} \\
D_{\mu } &=& \partial _{\mu } \; + \; i A_{\mu }
\nonumber \\
F_{\mu \nu } & = & \partial _{\mu } A_{\nu } \; - \; \partial _{\nu }
A_{\mu } \; ,
\nonumber
\ena
where $\lambda $ is the bare coupling constant of the scalar field,
$e_{0}$ is the bare coupling of the scalar fields to the gauge fields
$A_{\mu }$ and $T$ is the temperature in units of Boltzmann's constant.
The functional integration in (\ref{eq:1}) runs over all scalar modes
satisfying periodic boundary conditions in the Euclidean time direction.
$F_{\mu \nu }$ is the field strength tensor and $j$ is an external source
which couples linearly to the composite field $\phi ^{\dagger } \phi $.
The effective action for the order parameter $\phi ^{\dagger } \phi $
is defined by the Legendre transformation, i.e.,
\be
\Gamma [ (\phi ^{\dagger } \phi )_{c} ] \; = \; - \ln Z[j]
\; + \; \int d^{4}x \; (\phi ^{\dagger } \phi )(x) j(x) \; ,
\hbo
(\phi ^{\dagger } \phi )_{c}(x) \; := \;
\frac{ \delta \ln Z[j] }{ \delta j(x) } \; ,
\label{eq:2}
\en
and the effective potential $U(\phi ^{\dagger } \phi)$ is obtained by
inserting constant classical fields into the effective action $\Gamma
$.  We wish to consider the effect of a constant, external magnetic
field $H$, which is generated by the gauge fields
\be
A_{\mu }(x) \; = \; H \; x_{1} \delta _{\mu 2 }
\label{eq:3}
\en
in a certain gauge. Since the action in (\ref{eq:1}) is gauge invariant
and since our non-perturbative approach is gauge invariant
(in fact we will introduce the gauge invariant scalar condensate
as collective variable), any other gauge than that in (\ref{eq:3})
gives the same results.

The non-perturbative approach is given by a linearisation of the
$\phi ^{4}$-interaction by means of an auxiliary field, i.e.,
\bea
Z[j] &=&  \int {\cal D} \phi \; {\cal D} \phi ^{\dagger } \;
{\cal D } \chi \; \exp \bigl[ - \int d^{4}x  \bigl(
\frac{1}{2} (D_{\mu } \phi )^{\dagger } D_{\mu } \phi
\; + \; \frac{1}{4 e_{0}} F_{\mu \nu } F_{\mu \nu }
\label{eq:4} \\
&+& \frac{6}{\lambda } \chi ^{2} + [ \frac{1}{2} m^{2} - j(x) - i \chi ]
\phi ^{\dagger } \phi \;  \bigr) \bigr] \; .
\nonumber
\ena
After integrating out the scalar fields, we calculate the generating
functional $Z[j]$ by making a loop expansion with respect to the field
$\chi $ around its mean field value~\cite{la1,lo93}. The lowest
order of this expansion coincides with the leading order of
a $1/N$-expansion~\cite{lo93}, i.e.,
\bea
-\ln Z_{c}[j](T,H) & = & \int d^{4}x \;
\bigl( - \frac{3}{2 \lambda } ( M - m^{2} + 2 j )^{2}
\; + \; \frac{1}{ 4 e_{0} } F_{\mu \nu } F_{\mu \nu } \bigr)
\label{eq:5} \\
&+&  \Tr \ln [ -\partial ^{2} +
i ( \stackrel{ \leftharpoonup }{ \partial } _{\mu }
- \stackrel{ \rightharpoonup}{\partial } _{\mu } )
A_{\mu } + A_{\mu }^{2} + M ] \; ,
\nonumber
\ena
where $M$ is related to the mean field value $\chi _{0}$
by $\chi _{0} = \frac{i}{2} ( M - m^{2} - 2 j)$.
In practice, it is easier to obtain $M$ from the master field equation
\be
\frac{ \delta \ln Z[j] }{ \delta M } \; = \; 0 \; .
\label{eq:6}
\en
The trace in (\ref{eq:5}) extends over all modes satisfying
periodic boundary conditions in Euclidean space-time.
For a constant magnetic field in the gauge in (\ref{eq:3}),
the eigenvalues $\omega $ of the operator entering the logarithm
in (\ref{eq:5}) and their degeneracy $d$ can be easily calculated
(see e.g. \cite{ki86}), i.e.,
\be
\omega _{n, \gamma , k_{3} } \; = \;
(2 \pi T) ^{2} n^{2} \, + \, k_{3}^{2} \, + \,
(2 \gamma +1) H \, + \, M \; . \hbo
d_{n, \gamma, k_{3}} \; = \; \frac{ H a^{2} }{ 2 \pi } \; ,
\label{eq:7}
\en
where $\gamma = 0,1, \ldots , \infty $ labels the Landau levels,
$k_{3}$ labels the momentum parallel to the magnetic field, and the system has
been put in a large three dimensional box with side length $a$
for level counting reasons.
The loop contribution in (\ref{eq:5}) is, in Schwinger's proper
time regularisation,
\be
L \; = \; - \int _{1/ \Lambda ^{2} }^{\infty } \frac{ ds }{s}
\sum _{n =- \infty }^{\infty } \int \frac{ d k_{3} }{ 2 \pi }
a \; \frac{ H a^{2} }{ 2 \pi } \sum _{\gamma =0 }^{\infty }
\exp \bigl[ -s ( k_{3}^{2} + M + (2\pi T)^{2} n^{2} +
(2 \gamma +1 )H ) \bigr] \; ,
\label{eq:8}
\en
where $\Lambda $ is the proper time cutoff.

Before we calculate the effective potential, we would like
to rewrite the loop contribution $L$ as a bosonic partition function
in order to compare with the real time formulation of temperature
dependent field theory. To this end we recast $L$ in a more compact
form, i.e.,
\bea
L &=&  - \int ^{\infty }_{1/\Lambda ^{2} } \frac{ds}{s} \; \Tr _{3} \,
\sum _{n} \exp \{ -s [ (2\pi T)^{2} n^{2} + E^{2} ] \} \; ,
\label{eq:9} \\
E &=& \sqrt{ k_{3}^{2} \, + \, (2 \gamma +1) H \, + \, M } \; ,
\nonumber
\ena
where the trace $\Tr _{3}$ represents the summation over $k_{3}$ and the
Landau levels. If we take the derivative of $L$ with respect to $M$,
the $s$-integration can be performed, i.e.,
\be
\frac{ \partial L }{ \partial M } \; = \; \Tr_{3}^{(R)} \sum_{n}
\frac{ 1 }{ (2 \pi T)^{2} n^{2} \, + \, E^{2} } \; ,
\label{eq:10}
\en
where we have introduced a regularisation procedure for the trace
over the spatial degrees of freedom implying that we can safely take
the limit $\Lambda \rightarrow \infty $.
Using the identity
\be
\sum _{n=-\infty }^{\infty } \frac{1}{ n^{2} + a^{2} } \; = \;
\frac{ \pi }{a} \hbox{ctgh} ( \pi a ) \; ,
\label{eq:11}
\en
which is easily proven by noting that both sides in (\ref{eq:11})
have the same pole structure, we obtain
\be
\frac{ \partial L }{ \partial M } \; = \; \frac{1}{2T} \; \Tr_{3}^{(R)} \;
\frac{1}{ E } \, \hbox{ctgh} \frac{E}{2T} \; .
\label{eq:12}
\en
Integration of (\ref{eq:12}) with respect to M is straightforward, and
the result is (up to an unimportant constant)
\be
L \; = \; \Tr_{3} \, 2 \, \ln [ 1 - e^{- \frac{E}{T} } ]
\; + \; \frac{1}{T} \Tr_{3}^{(R)} \, E \; .
\label{eq:13}
\en
The first term of (\ref{eq:13}) is temperature dependent and finite
implying that the regularisation prescription can be removed.
It is just the partition function of two bosonic degrees of freedom
in an external magnetic field. The second term in (\ref{eq:13})
is divergent and (up to the trivial factor $1/T$) temperature
independent. It represents the mode sum of the zero point fluctuations
and is related to the Casimir effect (see e.g.\ \cite{la3}).

\medskip
{\it 3.\ The effective potential of the order parameter
$\phi ^{\dagger } \phi $ }

Since we have established, using (\ref{eq:13}), that we have not
missed any of the physical effects of a scalar particle in a magnetic
field by using the imaginary time formalism, we may proceed further by
starting with (\ref{eq:8}), which proves more convenient.  The
summations in (\ref{eq:8}) over $k_{3}$ and the Landau levels can be
performed and after a Poisson re-summation of the $n$-sum we obtain
\be
L \; = \; - \frac{ H a^{3} }{ 8 \pi ^{2} T }
\int _{1/ \Lambda ^{2} }^{\infty } \frac{ ds }{ s^{2} } \;
e^{-sM} \; \frac{ e^{-s H} }{ 1 - e ^{-2 s H} } \; \bigl[ 1 +
\sum _{\nu \not= 0} \exp ( - \frac{ \nu ^{2} }{ 4 s T^{2} } ) \bigr] \; .
\label{eq:14}
\en
Note that the divergences arise from the $s$-integration
at small $s$ values. This implies that only terms associated with
the $1$ in the square brackets in (\ref{eq:14}) contain divergences.
We split the loop $L$ into a finite part and a part containing
divergences, i.e.,
\bea
L &=& \frac{ a^{3} }{ 8 \pi ^{2} T } (L_{div} \; + \; L_{fin} )
\label{eq:15} \\
L_{div} &=& - H \int \frac{ ds }{ s^{2} } \;
e^{-sM} \, [ \frac{ 1 }{ 2 H s } \, - \, \frac{ H }{12} s
e^{-sH} ]
\label{eq:16} \\
L_{fin} &=& - H \int \frac{ ds }{ s^{2} } \;
e^{-s M } \, [ \frac{ e^{-sH} }{ 1 - e^{-2sH} } - \frac{1}{2Hs}
+ \frac{ H }{12} s e^{-sH} ]
\label{eq:17} \\
&-& H \int \frac{ ds }{ s^{2} } \;
e^{-sM} \, \frac{ e^{-s H } }{ 1 - e^{-2sH} } \, \sum _{\nu \not= 0}
\exp ( - \frac{ \nu ^{2} }{ 4 s T^{2} } ) \; .
\nonumber
\ena
The term $\frac{H}{12}s$ instead of the term $\frac{ H}{12} s e^{ -sH }$
in the square brackets in (\ref{eq:16}) would also cancel the
logarithmic ultra-violet divergence. However, the term linear in $s$
would cause a infra-red divergence in the limit $M \rightarrow 0$,
since the $s$-integral would diverge logarithmically at the upper
bound $s \rightarrow \infty $. The additional exponential factor does not
affect the ultra-violet behaviour, and implies that the counter term
does not influence the infra-red regime of the $s$-integration.
The ambiguity in defining the subtraction of divergences
is cured by renormalisation group invariance as usual.
Talking the limit $H \rightarrow 0 $ in (\ref{eq:16})
$L_{div}$ coincides with the divergent part which we would have
obtained by doing the calculation without a magnetic field (compare
\cite{la1})
\footnote{ Since we are dealing with a scalar field with two
degrees of freedom, the result differs by factor of two from that
in \cite{la1}. }.
This confirms that the right degeneracy factor $d$ occurs
in (\ref{eq:7}) for the Landau levels.

The $s$-integration in (\ref{eq:16}) can be easily performed, i.e.,
\bea
L_{div} & = & - \frac{1}{2} M^{2} \,
\Gamma(-2 , \frac{ M }{ \Lambda ^{2} }) \; + \;
\frac{ H^{2} }{ 12 } \, \Gamma (0, \frac{ M+H }{ \Lambda ^{2} })
\label{eq:18} \\
& \rightarrow & \frac{1}{2}  M
\Lambda ^{2} + \frac{1}{4} M^{2} ( \ln \frac{M}{\Lambda ^{2}}
- \frac{3}{2} + \gamma) \, - \, \frac{ H^{2} }{ 12 } \ln
\frac{ M+H }{ \Lambda ^{2} } \} \; ,
\label{eq:19}
\ena
where $\Gamma $ is the incomplete $\Gamma $-function and we have used
the asymptotic form of the $\Gamma $-function to obtain (\ref{eq:19}).
Inserting the loop $L$ in (\ref{eq:15}) into the expression for
$- \ln Z[j]$ in (\ref{eq:5}) one observes that the divergences
of (\ref{eq:19}) can be absorbed into the bare parameters by setting
\bea
\frac{ 6 }{ \lambda } \; + \; \frac{1}{8 \pi ^{2} } \bigl(
\ln \frac{ \Lambda ^{2} }{ \mu ^{2} } \, - \, \gamma \, + \, 1 \bigr)
\; = \; \frac{ 6 }{ \lambda _{R} } \; ,
\label{eq:20} \\
\frac{6}{\lambda } j \; - \; \frac{ 3 m^{2} }{\lambda } \; - \;
\frac{1}{ 16 \pi ^{2} } \Lambda ^{2} \; = \; \frac{ 6 }{ \lambda _{R} }
j_{R} \; - \; \frac{ 3 m^{2}_{R} }{ \lambda _{R} } \; ,
\label{eq:21} \\
j \; - \; m^{2} \; = \; 0 \; ,
\label{eq:22} \\
\frac{1}{ e_{0} } \; + \; \frac{ 1 }{ 48 \pi ^{2} }
\ln \frac{ \Lambda ^{2} }{ \mu ^{2} } \; = \; \frac{1}{e_{R}} \; .
\label{eq:23}
\ena
Although the magnetic field is external, we need a renormalisation
of the electric charge (\ref{eq:23}),
since the dynamical scalar particle couples to the magnetic field.
Mass, coupling constant and field renormalisation (\ref{eq:20}-\ref{eq:22})
are not affected by the presence of a magnetic field or temperature.
We have also obtained a renormalisation group invariant result, because
a change of the renormalisation point $\mu $ can be compensated
by a change of the renormalised parameters.
We refer to \cite{la1,la2,la3} for further details, but note
that here we are dealing with a scalar field with two
degrees of freedom, implying that the scaling functions also differ
by a factor of two from those of the real scalar field investigated
in \cite{la1,la2,la3}.

The effective potential is obtained by performing the Legendre
transformation (\ref{eq:2}) with respect to a constant
renormalised source $j_{R}$. So we have
\be
(\phi ^{\dagger } \phi )_{c} \; = \; \frac{1}{V_{4}}
\frac{ \partial \, \ln Z[j_{R}] }{ \partial j_{R} } \; = \;
\frac{6}{\lambda_{R}} M \; ,
\label{eq:24}
\en
where $V_{4}$ is the space-time volume. Due to field renormalisation,
$M = \lambda _{R} \phi ^{\dagger } \phi /6 $ is renormalisation
group invariant and will be referred to as the physical condensate
from here on (see also \cite{la2,la3}).

In order to remove the renormalisation point $\mu $, we introduce
a physical scale, e.g.,\ the scalar condensate at zero temperature
and zero magnetic field. In the latter case $L_{fin}$ in (\ref{eq:15})
vanishes, and the effective potential is given by
\be
U(M) \; = \; \frac{1}{32 \pi ^{2}} M^{2} \bigl(
\ln \frac{M}{\mu ^{2} } - \frac{1}{2} \bigr) \; - \;
\frac{3}{2 \lambda _{R} } M^{2} \; .
\label{eq:25}
\en
It has a global minimum at a non-vanishing value $M_{0}$ which
is determined by
\be
\frac{1}{16 \pi ^{2} } \ln \frac{ M_{0} }{ \mu ^{2} }
- \frac{ 3 }{ \lambda _{R} } \; = \; 0 \; .
\label{eq:26}
\en
At zero temperature and zero magnetic field, the state with the lowest
vacuum energy density has a non-vanishing scalar condensate
and is therefore non-trivial.

After inserting (\ref{eq:15}) into (\ref{eq:5}), performing the
renormalisation (\ref{eq:20}-\ref{eq:23}) and the Legendre
transformation we eliminate the renormalisation point $\mu $
in favour of the zero temperature condensate $M_{0}$ and
finally obtain
\bea
32 \pi ^{2} U(M,T,H) &=& M^{2} \bigl( \ln \frac{M}{M_{0}} -
\frac{1}{2} \bigr) \; + \; \bigl( \frac{ 16 \pi ^{2} }{ e_{R}(M_{0}) }
\, - \, \frac{1}{3} \ln \frac{ M+H }{ M_{0} } \bigr) \, H^{2}
\label{eq:27} \\
&-& 4 H \int _{0}^{\infty } \frac{ ds }{ s^{2} } e^{-sM}
\bigl[ \frac{ e^{-sH} }{ 1 - e^{-2sH} } \, - \, \frac{1}{2Hs}
\, + \, \frac{Hs}{12} e^{-sH}
\label{eq:28} \\
&+& \frac{ e^{-sH} }{ 1 - e^{-2sH} } \sum _{\nu \not= 0 }
e^{- \frac{ \nu^{2} }{ 4 s T^{2} } } \, \bigr] \; .
\label{eq:29}
\ena
This is our final result. It includes temperature effects as well as
the influence of the magnetic field.
The first two lines (\ref{eq:27}, \ref{eq:28}) are the effective
potential in the presence of a magnetic field at zero temperature.
The last line (\ref{eq:29}) gives the correction incurred at
finite temperature.

\medskip
{ \it 4.\ Results and discussion }

{}From solid state physics we know that magnetic fields promote ordering
whilst high temperature tend to destroy order and this competitive
behaviour is also observed when we study the effective potential
(\ref{eq:27}-\ref{eq:29}) numerically.  Magnetic fields prefer the
non-trivial phase whereas high temperatures drive the system towards
the trivial ground state.  This fact can be observed in figure 1,
which shows the effective potential at fixed temperature and differing
magnetic field strengths.  Starting at some temperature which is large
enough to obtain the perturbative phase,
\footnote{ In this phase the global minimum of the effective potential
is attached at $M=0$ implying that there is no scalar condensate. }
increasing the magnetic field causes a first order phase transition to
the non-trivial phase (with a non-vanishing scalar condensate). Figure
2 shows the critical magnetic field which is needed to restore the
non-trivial ground state as
a function of temperature. If the magnetic field is large, the system
can sustain large temperatures before it undergoes the phase
transition to the perturbative vacuum.  In figure 3, the effective
potential at the transition point is shown for pairs of the critical
temperature $T_{c}$ and the critical magnetic field $H_{c}$.  It is
observed that the barrier, separating the perturbative from the
non-trivial ground state, becomes smaller for large values of $(T_{c},
H_{c})$. In figures 1 and 3, we have omitted a term proportional to
$H^{2}$ which is independent of $M$, since the magnitude of this term
is governed by the additional parameter $e_{R}$.

It was shown that the presence of magnetic fields does not qualitatively
alter the effective potential for the order parameter $\phi ^{\dagger }
\phi $. This result is consistent with models which assume that the
phase transition in the early universe can be effectively described by a
scalar field (inflaton) regarding the remaining degrees of freedom as
spectators. However, in order to verify this assumption one must
extend our considerations to different types of external degrees of
freedom.

At the electro-weak phase transition the magnetic fields, produced by
the mechanism proposed by T.\ Vachaspati~\cite{va91} and further
investigated by K.\ Enquist and P.\ Olesen in \cite{en93}, are of
order $T_{c}^{2}$ with $T_{c}$ being the transition temperature. Our
considerations suggest that their influence on the transition is not a
perturbative correction.  However, their net effect is to increase the
transition temperature and to weaken the barrier between the
perturbative and the non-trivial vacuum. The latter effect indicates
that the transition is hastened due to the presence of magnetic
fields, though one should really study the effective action (as in
\cite{fo93}) to address this question.  Nevertheless our calculations
suggest that the dynamics of the phase transition, e.g.,\ nucleation
of bubbles turning false vacuum in true vacuum~\cite{co77}, is
strongly affected by magnetic fields.  This fact might have some
influence on electro-weak baryongenesis since the baryon asymmetry in
the universe (BAU) is intrinsically related to the dynamics of the
phase transition~\cite{mc91}.  For quantitative investigations in the
context of the electroweak phase transition one should apply the
non-perturbative approach, presented in~\cite{la1,la2,la3} and briefly
discussed above, to the complete standard model of weak interactions.
This will be an interesting task for future work.

\bigskip
\leftline{\bf Acknowledgements: }

I thank H.\ Reinhardt for encouragement and support. I am also
grateful to P.\ Elmfors for helpful discussions and to
P.\ Olesen for interesting remarks. I am indebted to R.\ F.\ Langbein
for carefully reading this manuscript as well as for useful comments.

\medskip

\vfill \eject
\centerline{ \large Figure captions }

\bigskip
{\bf Figure 1: } The effective potential $U$ as a function of the scalar
condensate $M$ at fixed temperature $T=0.35\, \sqrt{M_{0}}$
for $H=0$ (solid line), $H=0.25 \, M_{0}$ (dotted line) and
$H=0.5 \, M_{0}$ (dashed line). $U$ and $M$
in units of the condensate $M_{0}$ at zero temperature and zero magnetic
field.

\vspace{1 true cm}
{\bf Figure 2: } The critical magnetic field $H_{c}$ as a function of the
temperature $T$. $H_{c}$ and $T$ in units of $M_{0}$.

\vspace{1 true cm}
{\bf Figure 3: } The effective potential $U$ as a function of $M$
at the phase transition for $T=0.32, \, H=0.08$ (solid line),
$T=0.4, \, H=0.67 $ (dotted line) and $T=0.5, \, H=1.33$ (dashed line).
$U, M, T, H$ in units of $M_{0}$.

%
%
%
%

\begin{thebibliography}{sch90}
\bibitem{gu80}{ A.\ H.\ Guth, Phys.Rev. D20(1980)30. }
\bibitem{li83}{ A.\ D.\ Linde, Phys.Lett. B129(1983)177. }
\bibitem{as82}{ A.\ Albrecht, P.\ J.\ Steinhardt,
   Phys.Rev.Lett. 48(1982)1220. }
\bibitem{pi84}{ S.-Y.\ Pi, Phys.Rev.Lett. 52(1984)1725. \\
   Q.\ Shafi, A.\ Vilenkin, Phys.Rev.Lett. 52(1984)691. }
\bibitem{hol84}{ R.\ Holman, P.\ Ramond, G.\ G.\ Ross,
   Phys.Lett. B137(1984)343. }
\bibitem{ko90}{ {\it see e.g. \/} E.\ W.\ Kolb, M.\ S.\ Turner, `The
   Early Universe', Addison-Wesley, Redwood City, 1990. }
\bibitem{gu83}{ A.\ H.\ Guth, E.\ J.\ Weinberg, Nucl.Phys. B212(1983)321. }
\bibitem{fo93}{ R.\ F.\ Langbein, K.\ Langfeld, H.\ Reinhardt,
   'Natural Slow-Roll Inflation', T\"ubingen preprint October 1993. }
\bibitem{la1}{ K.\ Langfeld, L.\ v.\ Smekal, H.\ Reinhardt,
   Phys.Lett. B308(1993)279. }
\bibitem{la2}{ K.\ Langfeld, L.\ v.\ Smekal, H.\ Reinhardt,
   Phys.Lett. B311(1993)207. }
\bibitem{la3}{ K.\ Langfeld, F.\ Schm\"user, H.\ Reinhardt, 'Casimir
   effect of strongly interacting scalar fields', hep-ph 9307258,
   UNIT\"U-THEP-8$/$1993. }
\bibitem{ab76}{ L.\ F.\ Abbot, J.\ S.\ Kang, H.\ J.\ Schnitzer,
   Phys.Rev. D13(1976)2212. }
\bibitem{ste84}{ M.\ Stevenson, Z.Phys. C24(1984)87, Phys.Rev.
   D32(1985)1389. \\
   M.\ Stevenson, R.\ Tarrach, Phys.Lett. B176(1986)436. }
\bibitem{ok87}{ Anna Okopinska, Phys.Rev. D35(1987)1835. }
\bibitem{va91}{ T.\ Vachaspati, Phys.Lett. B265(1991)258. }
\bibitem{en93}{ K.\ Enquist, P.\ Olesen, 'On Primordial Magnetic
   fields of Electroweak Origin', Niels Bohr Institute preprint
   NBI-HE-93-33. }
\bibitem{ruz88}{ A.\ A.\ Ruzmaikin, A.\ A.\ Shukurov, D.\ D.\
   Sokoloff, 'Magnetic Fields of Galaxies', (Kluwer, Dordrecht, 1988). }
\bibitem{lo93}{ L.\ v.\ Smekal, K.\ Langfeld, H.\ Reinhardt,
   'Scaling improved $1/N$-expansion of non-trivial $\phi ^{4}$-theory',
   university of T\"ubingen, preprint October 1993. }
\bibitem{ki86}{ C.\ Kittel, 'Introduction to Solid State Physics',
   sixth edition, John Wiley, New York 1986. }
\bibitem{co77}{ S.\ Coleman, Phys.Rev. D15(1977)2929. \\
   C.\ G.\ Callan, S.\ Coleman, Phys.Rev. D16(1977)1762. }
\bibitem{mc91}{ L.\ McLerran, M.\ Shaposhnikov, N.\ Turok,
   M.\ Voloshin, Phys.Lett. B256(1991)451. }

%
%

\end{thebibliography}
\end{document}